\documentstyle[prl,aps,floats,epsfig]{revtex}

\draft

\def\beq{\begin{equation}}
\def\eq{\end{equation}}
\def\eeq{\end{equation}}

\newcommand{\gsim}{\lower.7ex\hbox{$\;\stackrel{\textstyle>}{\sim}\;$}}
\newcommand{\lsim}{\lower.7ex\hbox{$\;\stackrel{\textstyle<}{\sim}\;$}}

\def\mesino{{\cal M}}
\def\sbaryon{{\cal B}}
\def\sq{\widetilde Q}

\def\gl{\tilde g}
\def\wi{\widetilde W}

\def\qbar{\overline q}
\def\msq{m_{\sq}}

\def\mgl{m_{\gl}}

\def\eV{{\rm\,eV}}
\def\MeV{{\rm\,MeV}}
\def\GeV{{\rm\,GeV}}

\def\units#1{{\rm\,#1}}

\def\PRD#1#2#3{Phys.~Rev.~D {\bf #1}, #2 (#3)}

\def\NPB#1#2#3{Nucl.~Phys.~{\bf B#1}, #2 (#3)}

\newcommand{\met}{\rlap{\,/}E_T}

\begin{document}

\twocolumn[

\hsize\textwidth\columnwidth\hsize\csname @twocolumnfalse\endcsname

\title{Mesino -- Antimesino Oscillations}

\author{Uri Sarid$^a$ and Scott Thomas$^b$}
\address{{}$^{(a)}$Department of Physics, 
University of Notre Dame, Notre Dame, IN
46556 USA}


\address{{}$^{(b)}$Department of Physics, 
Stanford University, Stanford, CA 94305
USA}


\maketitle

\begin{abstract} 
The phenomenological implications of supersymmetric theories
with low scale supersymmetry breaking and a squark as the 
lightest standard model superpartner are investigated. 
Such squarks hadronize with light quarks, forming sbaryons 
and mesinos before decaying.  Production of these supersymmetric 
bound states at a high energy collider can lead to displaced jets 
with large negative impact parameter.  Neutral mesino--antimesino 
oscillations are not forbidden by any symmetry and can occur at 
observable rates with distinctive signatures. 
Stop mesino--antimesino oscillations would give a sensitive 
probe of up-type sflavor violation in the squark sector, 
and can provide a discovery
channel for supersymmetry through events with a same-sign top-top 
topology. 


\end{abstract}

\pacs{PACS numbers:  14.80.Ly, 12.60.Jv\hfill UND-HEP-98-US01 
\hfill SU-ITP-98-70 \hfill hep-ph/9909349}

]

Supersymmetry (SUSY) allows the standard model to be embedded
in more fundamental 
theories characterized by much higher energy scales, such as
the Grand Unified
or Planck scales, by stabilizing the electroweak scale against
quadratic radiative corrections. 
In order to maintain this stability, 
the supersymmetry breaking
masses for the superpartners of the ordinary particles
should be of order the electroweak scale. 
However, the underlying mechanism and scale 
of this spontaneous breaking, as well as the 
superpartner spectrum, remain at present largely unknown. 

If the intrinsic scale of spontaneous 
supersymmetry breaking is below a scale intermediate
between the electroweak and Planck scales, 
the lightest supersymmetric particle (LSP)
is the gravitino superpartner of the graviton.
The identity of the
next-to-lightest supersymmetric particle (NLSP) is model-dependent,
and largely determines the phenomenological signatures for
supersymmetry \cite{goldstino}.

In this letter the consequences of a squark NLSP are explored. 
While perhaps 
unconventional, a squark NLSP can occur in theories of gauge-mediated
supersymmetry breaking in which the strongly interacting 
messenger fields have suppressed couplings to spontaneous supersymmetry
breaking. 
Beyond the theoretical motivation, this scenario leads to very 
novel phenomenology. 
Squarks hadronize in sbaryon or mesino bound states.
Decay of these states can give rise to displaced jets
with angular distributions which differ significantly 
from displaced jets from heavy quark decay. 
In addition, neutral mesino--antimesino oscillations are not forbidden 
and could provide a very sensitive probe of sflavor violation
in the squark sector. 
Rather than representing a very specialized measurement
which could only be attempted once the existence of superpartners is
established,
mesino--antimesino oscillations can in fact 
provide a {\it discovery} channel for SUSY at high 
energy colliders. 


{\it Squark Decay and Hadronization}.---
With low scale supersymmetry breaking and 
conserved $R$-parity, the NLSP is metastable. 
The only available coupling for decay is through the 
Goldstino components, 
$\tilde G$, of the gravitino.
The two-body decay rate for an up-type squark NLSP, 
$\widetilde Q \rightarrow q \widetilde G$, including general 
(s)flavor misalignment between the quark and squark 
mass eigenstates, is given by 
\beq
\Gamma( \widetilde{Q}_a \rightarrow q_i \widetilde{G} ) =
{ m_{\tilde{Q}_a}^5 \over 16 \pi F^2}
  ~\epsilon_{(N)ai}^2
  \left( 1 - { m_{q_i}^2 \over  m_{\widetilde{Q}_a}^2 } \right)^4
  \label{tbgoldstino}
\eq
where $\sqrt{F}$ is the supersymmetry breaking scale, 
and 
$\epsilon_{(N)ai}^2 = 
  \epsilon_{(L)ai}^2 + 
  \epsilon_{(R)ai}^2 =
                      |\widetilde{U}_L U_L^{\dagger}|^2_{ai} + 
                      |\widetilde{U}_R U_R^{\dagger}|^2_{ai} $.
The left- and right-handed up-type quark
flavor-mass eigenstates are related to the gauge eigenstates by  
$q_{Li} = U_{Lij} q^{\prime}_{Lj}$ and 
$q_{Ri} = U_{Rij} q^{\prime}_{Lj}$ where $i,j=1,2,3$, 
and the up-type squark mass eigenstates are related to flavor 
eigenstates by 
$Q_a = \widetilde{U}_{Lai} \widetilde{Q}_{Li} + 
 \widetilde{U}_{Rai} \widetilde{Q}_{Ri} $ where $a=1,\dots,6$,
and likewise for the down-type (s)quarks. 

The lightest squark is likely to be mostly stop-like because
of left-right stop level repulsion, and
negative renormalization group evolution contribution to the 
mass, both proportional to the top Yukawa. 
For a stop-like squark lighter than the top quark, 
the only kinematically allowed 
two-body decays are through flavor violating suppressed
modes to light quarks, such as 
$\tilde t \rightarrow c \widetilde G$.
Three body decay through a charged current interaction, 
$\tilde t \rightarrow bW \widetilde G$,
is not suppressed by flavor violation, but 
is phase space suppressed.
The three-body decay rate in the limit in which the charginos and
all squarks expect the NLSP are heavy is given by 
$$
\Gamma(  \widetilde{Q}_a \rightarrow q_i W\widetilde{G} ) =
{ \alpha_2 m_{\tilde{Q}_a}^5 \over 128 \pi^2 F^2}
  \left[  
          \left| \sum_j \epsilon^{(q_j)}_{(L)ai} \right|^2
          ~I\left( {m_W^2 \over m_{\widetilde{Q}_a}^2} ,
                      {m_{q_j}^2 \over m_{\widetilde{Q}_a}^2} \right)
                  \right.
$$
\beq
  \left.
     ~~~~~~~~~   +
	 \left| \sum_j \epsilon^{(q_j)}_{(R)ai} \right|^2
          ~J\left( {m_W^2 \over m_{\widetilde{Q}_a}^2} ,
                      {m_{q_j}^2 \over m_{\widetilde{Q}_a}^2} \right)
  \right]
\eq
where 
$\epsilon^{(q_j)}_{(L)ai} = ( \widetilde{U}_L U_L^{\dagger} )_{aj}
    ( {U}_L D_L^{\dagger} )_{ji}$ and 
$\epsilon^{(q_j)}_{(R)ai} = ( \widetilde{U}_R U_R^{\dagger} )_{aj}
    ( {U}_L^{\dagger} D_L )_{ji}$ and 
$({U}_L^{\dagger} D_L )_{ji}=V_{ji}$ is the CKM quark mixing matrix. 
The phase space integrals are
$$
I(a,b) = \int_a^1 dx
   { (1-x)^4 (x-a)^2 \over 12 x^3 a}
     \left(
           {    6x^3(3a+x) \over (x-b)^2}
    \right.
$$
$$
   \left.
    ~~~~ + { 4x^2 (4a-x) \over (x-b) }
         + x^2 + 2xa + 3a^2
           \right)
$$
\beq
J(a,b) = \int_a^1 dx
    { b (1-x)^4(x-a)^2(2a+x) \over 2 x^2 a (x-b)^2}
\eq
For a stop-like squark 
the three-body mode $\tilde t \rightarrow bW \widetilde G$ 
dominates
if the sflavor violation is very small, 
while the two-body mode(s)
$\tilde t \rightarrow q_i \widetilde G$, $i=1,2$, 
dominate for larger sflavor violation. 
Numerically, for $m_{\tilde t} = 150$ GeV the three-body decay 
$\tilde t_R \rightarrow bW \widetilde G$ dominates for 
$\epsilon_{(R)i3} 
  \lsim 4 \times 10^{-3}$.
For a stop-like squark much heavier than the top, 
$\tilde t \rightarrow t \widetilde G$ dominates
$\tilde t \rightarrow q_i \widetilde G$ for 
$\epsilon_{(N)i3} \lsim (1 - m_t^2/m_{\tilde t}^2)^2$. 

For either the two- or three-body modes, decay of 
a squark NLSP to the Goldstino can take place over 
macroscopic 
distances \cite{goldstino},
and easily exceeds the hadronization 
length scale \cite{metanote}.
For example, with $m_{\tilde t} = 150$ GeV 
$\Gamma^{-1}(\tilde{t}_R \rightarrow bW \widetilde G) \simeq 
75~{\rm cm}~(\sqrt{F} / 100~{\rm TeV})^4$, while for 
$m_{\tilde t} = 190$ GeV
$\Gamma^{-1}(\tilde{t} \rightarrow t \widetilde G) \simeq 
0.75~{\rm cm}~(\sqrt{F} / 100~{\rm TeV})^4$.
A NLSP squark therefore always hadronizes before decaying.
Hadronization with a light antiquark leads to a 
neutral or charged mesino
bound state, 
$\mesino_{\widetilde Q \qbar} \equiv (\widetilde Q \qbar)$, 
while hadronization with two light quarks leads to a 
neutral or charged sbaryon 
bound state 
$\sbaryon_{\widetilde Q qq} \equiv (\widetilde Q qq)$, and likewise
for the antiparticle states. 

A long lived mesino or sbaryon might provide an interesting system
in which to study heavy (s)quark aspects of QCD. 
In this work we concentrate on the implications for 
supersymmetric and sflavor phenomenology. 


{\it Mesino Oscillations}.--
The mesino bound states are spin ${1 \over 2}$ Dirac fermions.
A neutral mesino and antimesino differ by 
two units of (s)flavor, fermion number, $F$, and $R$-charge. 
All of these quantum numbers are, however, manifestly violated in 
any supersymmetric theory.
(S)quark flavor is violated by Yukawa couplings
and squark flavor may also be violated by scalar tri-linear
couplings and possibly the scalar mass-squared matrices. 
Fermion number and $R$-symmetry are violated by gaugino 
masses. 
So hadronization of squarks into neutral mesino bound states
allows for the interesting phenomenon of particle--antiparticle
oscillations which is impossible for an isolated charged particle. 
Operators which violate the above symmetries 
appear as Majorana mass terms which mix the mesino and 
antimesino states. 
Including these effects, mesinos are therefore 
pseudo-Dirac fermions.

Mesino oscillations are analogous to meson oscillations. 
At the microscopic level the
$\Delta \tilde Q = \Delta q = \Delta F = \Delta R =2$
amplitudes which mix mesino and antimesino 
arise from tree-level gluino and neutralino
exchange as illustrated in Fig. 1(a) for 
$\mesino_{\tilde{u}_a \bar{u}_i} \leftrightarrow
\overline{\mesino}_{\tilde{u}_a^* {u}_i}$. 
The tree level gluino contribution to the oscillation amplitude 
for a single linear combination 
of the Weyl components of the light quark in the color singlet 
channel is 
\beq
 {\cal A}_{\tilde g}
    (\widetilde{Q}_a \overline{q}_i  \rightarrow \widetilde{Q}_a^* q_i ) =
  ~ \epsilon_{(N)ai}^2   ~
   \left(1 - {1 \over N_c^2} \right)
         ~{ g_{s}^2 m_{\tilde{g}} \over
               m_{\tilde{g}}^2 - m_{\widetilde{Q}_a}^2 }
\eq
where
$\epsilon_{(N)ai}$ is the same flavor matrix which appears
in the two-body decay (\ref{tbgoldstino}), 
and $N_c=3$ is the number of colors.
Neutralino contributions are comparable depending on the precise spectrum.

\begin{figure}[ht]
\begin{center}
\epsfig{file=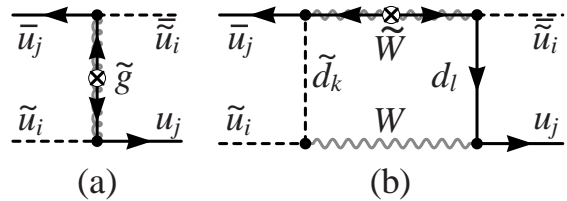,width=2.9in}
\vskip 10pt
\caption{
Mesino oscillations arise (a) from tree-level 
gaugino exchange and (b) from one-loop charged current--chargino
box diagrams. 
Arrows indicate the flow of fermion number and crosses represent
fermion number violating gaugino mass insertions. 
(S)flavor violation may be isolated in the gaugino and gauge
interactions. 
Diagrams for other neutralinos and charginos 
and those related by crossing are 
not shown.} 
\label{fig:osc} 
\end{center} 
\end{figure}

One-loop box diagrams also allow  
charged current and chargino contributions as illustrated in 
Fig. 1(b). 
In the pure $SU(2)_L$ Wino limit,
and ignoring left-right mixing in couplings for simplicity,
the one-loop amplitude is 
$$
 {\cal A}_{\widetilde W} 
   (\widetilde{Q}_a \overline{q}_i  \rightarrow \widetilde{Q}_a^* q_i) =
 { 8 \alpha_2^2 m_{\widetilde{W}} \over 3 m_{\widetilde{Q}_a}^2  }
 \sum_{b,\ell}
 \epsilon_{(C)ia}^{(\widetilde{Q}_b)}
 \epsilon_{(C)ia}^{(q_{\ell})}
$$
\beq
\times~ f( m_{\widetilde{Q}_b}^2/m_{\widetilde{Q}_a}^2  ,
    m_{{q}_{\ell}}^2/m_{\tilde{Q}_a}^2  ,
        m_{\widetilde{W}}^2/m_{\widetilde{Q}_a}^2  ,
    m_{{W}}^2/m_{\widetilde{Q}_a}^2   )
\eq
where 
$\epsilon_{(C)ia}^{(\tilde d_b)} = 
(U^\dag_L \widetilde D_L )_{ib}
(\widetilde D^\dag_L \widetilde U_L )_{ba}$ 
and
$\epsilon_{(C)ia}^{(d_\ell)} =
(U^\dag_L D_L)_{i\ell}
(D^\dag_L \widetilde U_L )_{\ell a} $
and the loop function $f(a,b,c,d)$ is 
\beq
\int_0^1 dx  \int_0^x  dy  \int_0^y  dz
                \left(   {(1-y^2) \over [D(a,b,c,d)]^2 } +
                   { 2 \over D(a,b,c,d) }  \right)
\eq
where
$D(a,b,c,d)= y^2 + (b-c-1)y + (c-a)x + (d-b)z + a$.
Ignoring quark masses 
$\sum_{\ell} \epsilon_{ia}^{(d_{\ell})} = 
\epsilon_{(L)ia}$.
The box contribution is intrinsically suppressed 
compared with the tree by 
at least $10^{-3}$
even without significant squark GIM suppression. 
So for up-type squarks the tree-level sflavor violating contributions
through gluino exchange dominate unless up-type sflavor violation 
is highly suppressed compared with down-type. 
Quark flavor mixing effects in charged current interactions
represent an irreducible contribution to mesino oscillations
through the box diagram, but are GIM suppressed by 
${\cal O}(m_b^2/m_{\widetilde Q}^2)$ and numerically unimportant
for oscillations which could be observed in flight. 

The mesino--antimesino oscillation frequency is related to the 
short distance amplitudes given above by 
\begin{equation}
\omega = \Delta m_{\mesino \overline \mesino} = 
{N_c \over 2\msq}  |\psi(0)|^2 
  \left|  
{\cal A}(\widetilde{Q}_a \overline{q}_i 
   \rightarrow \widetilde{Q}_a^* q_i )  
   \right|
\label{eq:mMMdef}
\end{equation}
The light (anti)quark wave function at the origin in the 
(anti)mesino is defined here in terms of the 
matrix element of the Dirac scalar bilinear,
$ |\psi(0)|^2 \equiv (\langle \mesino | \overline q q \delta^3(r) 
  | \mesino \rangle
 /  \langle \mesino | \mesino  \rangle) V$,
where $V$ is the normalization volume.
This may be related 
to the mesino decay constant by 
 $|\psi(0)|^2 = f_M^2 \msq/(4 N_c)$. 
In the heavy
(s)quark limit, this is a purely QCD quantity which may be approximated
using the $B$-meson 
decay constant $f_B \simeq 160$ MeV
giving $|\psi(0)|^2 \simeq (220\MeV)^3$ \cite{ref:Bparam}, 
independent of 
$\msq$ and $N_c$ in the heavy (s)quark and large $N_c$ limits. 

Numerically, the gluino contribution to the
$\mesino_{\tilde{t} \overline u} \leftrightarrow
\overline{\mesino}_{\tilde{t}^* u}$ 
oscillation wavelength is 
$\beta \gamma \lambda$ where 
$\lambda \equiv 2 \pi / \omega$ is 
\begin{equation}
\lambda \simeq (\mbox{4 nm}) \,
\left(\mgl\over250\GeV\right)^2 y\,(1-y^2)
~\epsilon_{(N)13}^{-2}
\label{eq:osclength}
\end{equation}
where $y=m_{\tilde{t}} / m_{\tilde g}$ and 
by assumption $y < 1$ so 
$0 < y(1-y^2) < \sqrt[3]{2/3} \simeq 0.38$.
Oscillations on the scale of a detector could occur 
for $\epsilon_{(N)13}$ as small as $5 \times 10^{-5}$. 
Up-type sflavor violation involving the third generation is 
essentially 
unconstrained by present data, so extremely rapid oscillations
compared with the decay length and 
scale of a detector are conceivable.  

The time integrated probability for a mesino to 
decay as an antimesino depends on the oscillation frequency
and decay rate
\beq
{\cal P}( \mesino \rightarrow \overline \mesino) = 
{x^2 \over 2(1 + \delta^2 + x^2) }
\label{PMM}
\eq
where $x = \Delta m_{\mesino \overline \mesino} / \Gamma$, 
and $\delta =  (m_{\mesino}-m_{\overline \mesino})/\Gamma$ 
allows for a possible diagonal flavor-conserving splitting
discussed below.
Rapid oscillations, 
$x \gg 1$, 
yield ${\cal P}( \mesino \rightarrow \overline \mesino) 
   = \frac12$,
while for slow oscillations, $x \ll 1$,  
${\cal P}( \mesino \rightarrow \overline \mesino) \to \frac12 x^2$.




{\it Experimental Signatures}.---
Since a squark NLSP is strongly interacting it is likely to 
be the most abundantly produced SUSY particle at a hadron collider. 
The experimental signatures depend on
the decay length. 
For a decay length of order or larger than a meter, non-relativistic 
squarks which hadronize as charged mesinos and sbaryons appear as 
highly-ionizing tracks (HITs) in a tracking chamber. 
Using the stop production cross section \cite{ref:stopprod}
and 
preliminary results of a CDF search for HITs 
in Run I at the Tevatron \cite{ref:stuart}
we estimate a current bound on the mass of long lived stops of 
roughly 150 GeV. 

Mesino or sbaryon 
decay lengths shorter than roughly a meter 
can lead to observably displaced jets with both large transverse
energy ($E_T$) and large missing transverse energy ($\met$).
For sufficiently long squark decay lengths, 
these may be distinguished from analogous heavy quark 
decays by large beam axis impact parameters. 
In addition, since the massive mesinos and sbaryons are 
non-relativistic, the decay
products are not significantly boosted in the lab 
frame and are therefore 
roughly uniformly distributed
in $\cos \varphi$, where $\varphi$ is the angle between the 
reconstructed momentum vector of the displaced jet and 
the unit normal between the beam
axis and the origin of the displaced jet. 
In contrast, the distribution of 
high $E_T$ displaced jets from direct heavy quark
production and  
decay is peaked near $\cos \varphi \sim 1$ since 
the visible decay products are boosted in a direction 
away from the production vertex. 
Mesino or sbaryon decay may therefore be distinguished by 
high $E_T$ jets with large $\met$ and 
with large negative impact parameters (LNIPs), 
$\cos \varphi \lsim 0$. 
These observables result if the visible decay 
products recoil against the invisible Goldstino  
in a direction towards, rather than away from the 
production vertex. 
LNIPs 
provide an efficient means to search for exotic massive 
metastable particles which decay to hadronic final 
states \cite{LNIPnote}.

Neutral 
mesino--antimesino oscillations present the possibility of another
novel experimental signature, even for decay lengths which are
too short to be resolved in real space. 
Oscillations may be revealed in any decay mode which tags
the sign of the (anti)squark in a neutral (anti)mesino. 
Squark--antisquark production 
events in combination with mesino--antimesino oscillation
can then lead to same-sign events. 
For example, the antisquark may hadronize as a neutral
antimesino which oscillates to a mesino before decaying, 
while the squark hadronizes as a charged mesino or sbaryon
which can not oscillate. 
Summing over all possibilities,
the time-integrated ratio of same- to
opposite-sign events is
\begin{equation}
R \equiv {N_{++} + N_{--} \over N_{+-} + N_{-+}}
= {2 {\cal P} f_0 (1 - {\cal P} f_0) 
   \over 1 - 2 {\cal P} f_0 + 2 {\cal P}^2 f_0^2}
\simeq 2 {\cal P} f_0 + 2 {\cal P}^2 f_0^2\,
\label{eq:Ratio}
\end{equation}
where ${\cal P} \equiv {\cal P}(\mesino \rightarrow \overline \mesino)$,
and $f_0$ is the neutral mesino hadronization fraction 
(the strange quark finite mass implies 
$\frac13 \lsim f_0 \lsim \frac12$ for up-type squarks).
Thus, for $x \gsim 1$ a significant fraction of squark--antisquark
events will yield same-sign events. 

The feasibility of determining the sign of an (anti)squark 
at decay depends on the decay products. 
For stop-like squark decays 
$\tilde t \rightarrow b W \widetilde G$ or 
$\tilde t \rightarrow t \widetilde G$ with 
$t \rightarrow bW$, the $W$-bosons
reliably tag the sign of the (anti)squark. 
The $W$-bosons signs may in turn 
be determined with the leptonic decay mode 
$W \rightarrow \ell \nu$ where $\ell=e,\mu$.
This requires isolating these primary 
leptons from any secondary leptons arising
from $b$-quark decay. 
Such distinctive, essentially background free
events have the topology of same-sign top-top
events, and provide a possible discovery mode for SUSY. 
The largest background is probably from 
top-antitop production with the very small 
probability of misidentification of the primary 
leptons or mismeasurement of the charges. 
At the Fermilab Tevatron Run IIa with 2 fb$^{-1}$ of 
integrated luminosity,
a 175 GeV stop squark with 
dominant decay $\tilde t \rightarrow bW \widetilde G$ and 
oscillation parameter 
$x \sim 1$ would yield $\sim 10$ same-sign dilepton top-top
events, while $x \gg 1$ would yield $\sim 20$ events. 
A detection acceptance times efficiency $\gsim 30$\%
should give a detectable signal for these parameters. 

Observation of stop mesino oscillations requires, and 
very sensitively probes, up-type squark sflavor violation. 
For example, for a stop decay length $\Gamma^{-1} \sim 10$ cm, 
maximal  
$\mesino_{\tilde{t} \bar{u}} \leftrightarrow
\overline{\mesino}_{\tilde{t}^* {u}}$
mixing, $x \gsim 1$, occurs for all 
$\epsilon_{(N)13} \gtrsim
5 \times10^{-5}$.
Even for $\Gamma^{-1} \sim 2\units{\mu m}$
(which could not be resolved as a displaced vertex), 
maximal mixing occurs for any 
$\epsilon_{(N)13} \gtrsim 10^{-2}$.
The magnitude of squark sflavor violation depends on the 
scale at which (s)quark flavor is broken. 
If the flavor scale is not too much larger than the messenger
scale for transmitting supersymmetry breaking, interesting
levels of sflavor mixing are expected, and 
observable mesino oscillations can occur. 

The flavor violating two-body decay $\tilde t \rightarrow c 
\widetilde G$ can dominate if sflavor violation is large enough, 
as discussed above. 
This mode also dominates if the NLSP squark
is scharm-like, 
$\tilde c \rightarrow c \widetilde G$. 
Semi-leptonic decay of the $c$-quarks 
hadronized in $D^{0,\pm}$-mesons could
then be used to tag same-sign events 
in high $E_T$ 
charm-jets with large $\met$.
$D^0 \leftrightarrow \overline D^0$ oscillation is 
negligible and would not contaminate a mesino oscillation
signal at the discovery level in a relatively clean sample
of LNIPs.  
However, squark decays in this mode with a decay length that
is too short to resolve using LNIPs would be contaminated
by standard model production of 
$b$-jets which are not easily distinguished from charm-jets. 
This standard model background could be significant since 
$B^0 \leftrightarrow \overline B^0$ oscillations are non-negligible.
Self-tagging of the heavy flavor at production 
to determine its sign \cite{ref:selftag}, or measuring total jet charge
after decay to isolate $D^{\pm}$
mesons which do not oscillate
could reduce this background, but requires large
statistics, and is probably 
not applicable at the discovery level. 

Observing oscillations for 
a squark NLSP which decays to other flavors is more problematic. 
For a sbottom-like squark which decays by 
$\tilde b \rightarrow b \widetilde G$, the $B^0$-meson backgrounds
discussed above are important. 
Decays to lighter quarks, 
$\widetilde Q \rightarrow (u,d,s)\widetilde G$, are difficult
to sign using self tagging. 
So a stop-like NLSP squark provides the best opportunity to observe
mesino oscillations. 

Charge conjugate non-invariant environmental effects
give rise to flavor conserving diagonal 
mesino--antimesino splitting, 
indicated by $\delta \neq 0$ in (\ref{PMM}),
and could potentially
suppress oscillations. 
The mesino spin couples to the 
ambient magnetic field, $B$, and induces a diagonal splitting.
Using the chiral quark value for the mesino magnetic
moment, $\mu_{\mesino} \simeq {2 \over 3} \mu_p$, gives
$\delta \lambda \equiv (2 \pi / \omega)(x/\delta) \simeq 
10.5~{\rm m}/[B/{\rm Tesla}]$. 
Forward scattering off nuclei in 
matter through operators of the 
form $(4 \pi /\Lambda_{\chi})^2~\overline{\mesino} \gamma^{\mu} 
\mesino~\overline{N} \gamma_{\mu} N$ 
where $\Lambda_{\chi} \simeq 1.1$ GeV is the 
chiral symmetry breaking scale, 
give
$\delta \lambda \simeq 1.2~{\rm m}/[ \rho/({\rm gm}~{\rm cm}^{-3})]$. 
So for oscillations which could be observed in flight through
the (low density) tracking region of 
a detector, these effects do not dominate, 
$\delta \lambda \gsim \lambda$ or 
$\delta / x \lsim 1$ \cite{spositron}.


{\it Conclusions}.--
A squark NLSP can give rise to novel experimental signatures
including HITs or LNIPs. 
The possibility of mesino--antimesino oscillations provides 
a discovery channel for SUSY through same-sign 
dilepton events in association with high $E_T$ heavy 
flavor jets and large $\met$. 
Oscillations are most easily observed for a stop-like squark
in same-sign top-top events, or same sign charm-jets
in a sample of LNIPs. 
Mesino oscillations would provide a very sensitive probe
of sflavor violation. 
If the mesino decay length
is macroscopic {\it and} the oscillation length
is fortuitously of the same order,  
$x \sim 1$, mesino oscillations could be observed in real space
in the signed decay length distributions. 

We 
thank I. Bigi, R.~Demina, Y.~Grossman, and 
N.~Uraltsev for useful discussions. 
The work of U.S.\ was supported in part by the US National Science
Foundation under grant PHY98-02483, and that of S.T.\ by 
the US National Science Foundation under grant PHY98-70115, 
the Alfred P. Sloan Foundation, and Stanford
University through the Frederick E.~Terman Fellowship.


\begin{thebibliography}{99}
\bibitem{metanote}
Metastable squarks
which hadronize and decay over macroscopic distances can also arise
with suppressed but non-vanishing renormalizable $R$-parity violation.
\bibitem{spositron} Matter effects would however suppress 
analogous oscillations in the
case of a spositronium bound state of 
an electron and a NLSP antislepton stopped in matter.
\end{thebibliography}
\end{document}